\documentclass[12pt]{article}
\usepackage{graphicx}

\begin{document}

\begin{center}
{\Large Low-dimensional light-emitting transistor with tunable recombination zone}
\end{center}

\begin{center}
B. Kaestner$\dagger$, J. Wunderlich$\ddagger$, and T.J.B.M. Janssen$\dagger$\\
$\dagger$ National Physical Laboratory, Hampton Road, Teddington, TW11 0LW.\\ 
$\ddagger$ Hitachi Cambridge Laboratory, Madingley Road, Cambridge, CB3 0HE.
\end{center}

\begin{quote}
We present experimental and numerical studies of a light-emitting transistor comprising two quasi-lateral junctions between a two-dimensional electron and hole gas. These lithographically defined junctions are fabricated by etching of a modulation doped GaAs/AlGaAs heterostructure. In this device electrons and holes can be directed to the same area by drain and gate voltages, defining a recombination zone tunable in size and position. It could therefore provide an architecture for probing low-dimensional devices by analysing the emitted light of the recombination zone.
\end{quote}

\section{Introduction}

Electroluminescence (EL) generated in $p$-$n$ junctions is primarily used for illumination purposes, such as in lamps, displays or optical fiber communication systems. More recently, the photon statistics and polarization properties have been studied. For instance, sensitive measurements of the emitted light showed that the noise properties can be exploited\cite{giacobino}. In nanoscale $p$-$n$ junctions, light sources have been realized which emit single photons when triggered by an electrical gate\cite{kim1PAII}. In recent years $p$-$n$ junctions have also been developed into effective probing tools, such as spin sensitive light-emitting diodes to detect electrical spin injection from magnetic into non-magnetic semiconductors \cite{fiederling, ohno} and to observe spin-transport phenomena in low dimensions \cite{wunderlich1, kaestner4, kaestner5}. In acoustoelectric nanocircuits $p$-$n$ junctions may be used to read out position \cite{foden} and spin \cite{barnes} of single electrons for devices in quantum information processing. A suitable lateral $p$-$n$ junction for the latter application was demonstrated by Hosey {\itshape et al.} \cite{hosey}.

In most of these devices a lateral low-dimensional geometry of the $p$-$n$ junction would significantly enhance their performance \cite{cecchini} or is an essential requirement. The properties of these junctions include very small capacitances and the possibility to manipulate the two-dimensional electron flow by gates \cite{foden}. Optoelectronic devices based on lateral junctions have been demonstrated (for example in \cite{vaccaro, kaestner2PBII}). However, in a lateral geometry the recombination zone is not bound and its size depends on the bias which decreases the recombination efficiency and causes problems when applied as a detection tool. One possible approach might be the incorporation of a gate contact to confine electrons injected into the $p$-type region which has been studied numerically by Ryzhii {\itshape et al.} \cite{ryzhii}. One difficulty arises from the opposite ways in which electrons and holes respond to a gate close to a $p$-$n$ junction. In addition, leakage might be present as a biased Schottky contact will always be conductive for one of the two carrier types.

Here we present a lateral EL device with control over the recombination zone, comprising two quasi-lateral junctions between a two-dimensional electron and hole gas (2DEG and 2DHG). These lithographically defined junctions are fabricated by etching of a modulation doped GaAs/AlGaAs heterostructure introduced in \cite{kaestner1PBII}, and combined into a transistor structure. Using the four terminals of this device the size and position of the recombination zone can be adjusted.

\section{Device design}

\begin{figure}
 \centerline{
 \includegraphics[scale=0.6]{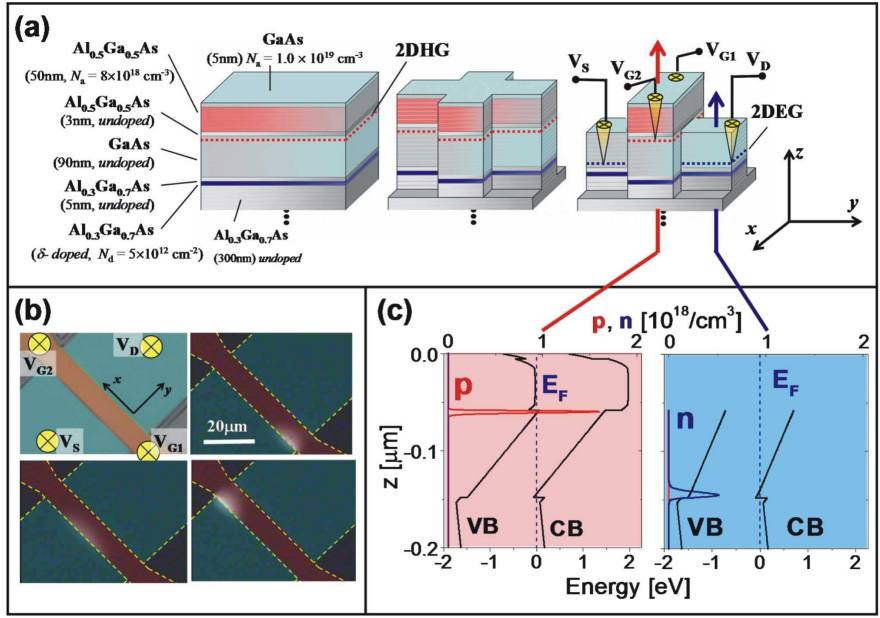}
 }
 \caption{Schematic diagram of the device, showing its composition and geometry in (a). The fabricated device is shown in (b), the hole containing channel is highlighted in yellow and is about $10\,\mu$m wide. The distribution of infrared light emission for different relative gate voltages $V_{G1}$ and $V_{G2}$ as recorded by a CCD camera is shown as a white area. The temperature was kept at 10$\,$K. In (c) the simulated band diagram as a function of $z$ is shown. Electron and hole densities are given by the blue and red curve, respectively. The blue and red arrow indicate the location on the device to which the band structure calculation belongs.}
 \label{Device}
\end{figure}

The wafer was grown on a (100) semi-insulating GaAs substrate and is shown schematically in Figure \ref{Device}(a). It consist of a Si-$\delta$-doped Al$_{0.3}$Ga$_{0.7}$As layer (sheet doping density $N_s = 5 \times 10^{12}$ cm$^{-2}$, spaced 5$\,$nm from the upper interface) of 300$\,$nm thickness was grown onto a semi-insulating GaAs substrate and subsequently covered with an undoped GaAs layer (90$\,$nm), an undoped Al$_{0.5}$Ga$_{0.5}$As spacer layer (3$\,$nm) and a layer of Be-doped Al$_{0.5}$Ga$_{0.5}$As (50$\,$nm, $N_a = 8 \times 10^{18}$ cm$^{-3}$). On top, a Be-doped GaAs capping layer was added. The layer structure was designed to accommodate two parallel conducting layers within the 90$\,$nm thick {\it i}-GaAs region: a hole gas at the upper interface and an electron gas at the lower interface. In the as-grown state, the electron gas is fully depleted such that the hole gas forms the only conducting layer, as shown by the red dotted line in Figure \ref{Device}(a). The structure is then patterned into a cross shaped mesa by etching below the Si-$\delta$-doped Al$_{0.3}$Ga$_{0.7}$As layer (centre of Figure \ref{Device}(a)). In the third stage, the upper, $p$-type material is removed for two arms of the cross-shaped mesa (right part of Figure \ref{Device}(a)). This lowers the electrostatic potential of the $i$-GaAs channel such that the previously depleted electron gas forms at the lower interface, indicated by the blue dotted line in Figure \ref{Device}(a). Thus, a junction between the electron and hole gas can be produced at the two step-edges. 

The structure is designed to be used at a temperature\footnote{The chosen temperature was determined by the application. A similar design is possible at room temperature, if required.} between 5 - 10 K and Figure \ref{Device}(c) shows the band diagram calculated along the $z$-axis at this temperature. The corresponding locations for the two distinct band diagrams in the actual device are indicated by the red and blue arrows in Figure \ref{Device}(a).

The measured device is shown in Figure \ref{Device}(b) and fabrication details can be found in \cite{kaestner3}. The hole containing channel is highlighted yellow and the corresponding contacts are indicated by the yellow circles. The structure can be operated in two different ways. If no voltages are applied, i.e. $V_{G1} = V_{G2} = V_D = V_S = 0$, the structure is in equilibrium such that the 2DHG-channel is continuous while the 2DEG channel contains a gap of about $10\,\mu$m. Applying a positive gate voltage $V_{G1}$ or $V_{G2}$ to the 2DHG, while keeping $V_D = V_S = 0$, populates electrons at the lower interface of the $i$-GaAs region, and therefore generates a channel between the two 2DEG regions. This operation is reminiscent of an $n$-channel field effect transistor (FET). For narrow 2DHG channel, applying a positive drain or source voltage $V_D$ or $V_S$ will deplete the 2DHG bordering the 2DEG. Here the same structure functions as a $p$-channel FET. This double nature of the structure can be used to gate both, electrons and holes, which is required to tune the recombination zone.

\section{Electroluminescence}

\begin{figure}
 \centerline{
  \includegraphics[scale=0.8]{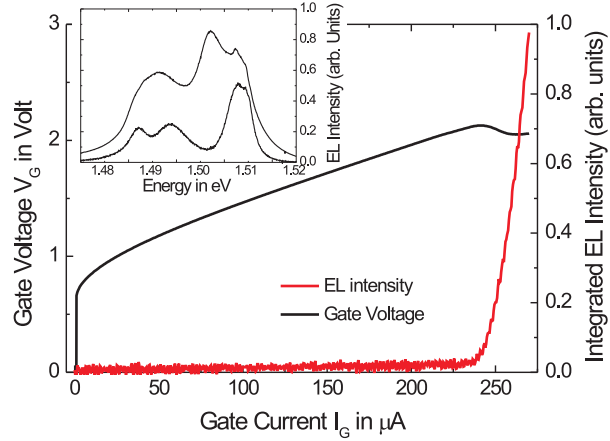}} 
 \caption{Required gate voltage (black) and EL intensity (red) as function of forward bias gate current $I_G$. Junction at drain is floating. Inset shows EL spectrum for two different gate currents $I_G$.}
 \label{EL}
\end{figure}

First we discuss the electrical conditions under which the device emits light. When applying a positive voltage $V_G = V_{G1} = V_{G2}$ with respect to $V_S = 0$, whereas $V_D$ is left floating, the device can be considered as a simple diode. To observe light emission the sample was mounted in a continuous flow optical cryostat. The temperature was kept at $T = 10\,$K. The current-voltage characteristic is shown in Figure \ref{EL}. As soon as a finite current $I_G$ through the gate was measured the required voltage $V_G$ had to be raised above $0.7$ V, which is below the band gap energy of GaAs of about $1.4\,$eV. As long as the current was kept smaller than $250\,\mu$A no light could be detected. The resulting current is believed to be caused by {\itshape electrons} flowing directly to the contact of $V_{G1}$ and $V_{G2}$. This is possible because $V_{G1} = V_{G2} > 0$ induces electrons at the lower interface of the $i$-GaAs region all the way to the contact of $V_{G1}$ and $V_{G2}$, which was confirmed by Hall measurements. Owing to the large built-in field in $z$-direction and a high impurity concentration underneath the $p$-ohmic contact, electrons may tunnel via impurity states towards the upper interface of the $i$-GaAs channel, where they recombine with holes.

As the current is increased further the resistances of the channel and tunneling processes will cause the radiative recombination across the 1.4$\,$eV bandgap to be more favorable. This switching between transport channels results in a negative differential two-terminal resistance, as shown in Figure \ref{EL}. At this point light emission starts suddenly, but its spectrum depends strongly on $V_G$ as shown in the inset of Figure \ref{EL}.

An important feature of this onset of light emission is that the recombination takes place in a {\itshape single} spot and is shifted towards the step-edge, as shown by the bright region in Figure \ref{Device}(b). Since the channel in $x$-direction is much longer than the $x$-dimension of the recombination zone one might expect two light spots on either side of the channel with their relative intensities determined by the relative voltages $V_{G1}$ and $V_{G2}$. The shift towards the step-edge may be understood by assuming that electrons continue to flow directly to the contact of $V_{G1}$ and $V_{G2}$, even when light emission has started and $I_G \ge 250\,\mu$A. This coexistence of a radiative and non-radiative channel causes a voltage drop which will only allow electrons near the step-edge to recombine radiatively across the bandgap. A simulation of the electrostatic potential distribution in the presence of two channels will be discussed in Section \ref{DevSimChap}.

\section{Tuning the recombination zone}

\begin{figure}
 \centerline{\includegraphics{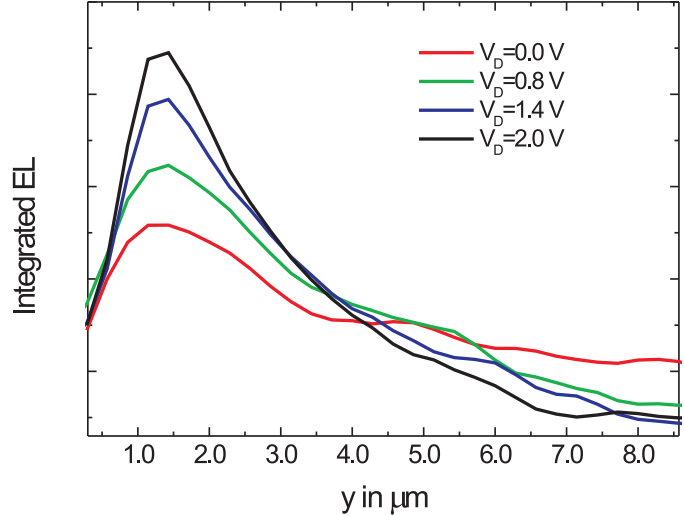}} 
 \caption{EL intensity integrated over $x$ along the $y$ direction for different values of the drain voltage $V_D$. As $V_D$ is increased the radiative recombination concentrates on the left hand side of the junction. The temperature was kept at around $10\,$K.}
 \label{CONF}
\end{figure}

By lowering the temperature the confinement of the recombination zone can be further enhanced which is required for probing applications \cite{nomura}. However, the confinement in the $y$-direction may also be controlled electrically by applying a voltage, $V_D$, to the drain contact. For a drain voltage $V_D$ larger than $V_{G1}$ and $V_{G2}$ both electrons and holes are repelled from the drain side as the junction between the 2DHG and the 2DEG is reverse biased. The junction on the source side, however, is forward biased as long as the source voltage $V_S = 0$ and $V_{G1}$ and $V_{G2}$ are positive. As a result the size of the recombination zone can be tuned by $V_D$ as shown in Figure \ref{CONF}. The light intensity along the $y$-direction integrated over $x$ is plotted for different drain voltages $V_D = 0, -, 2\,$V. 

\begin{figure}
 \centerline{\includegraphics{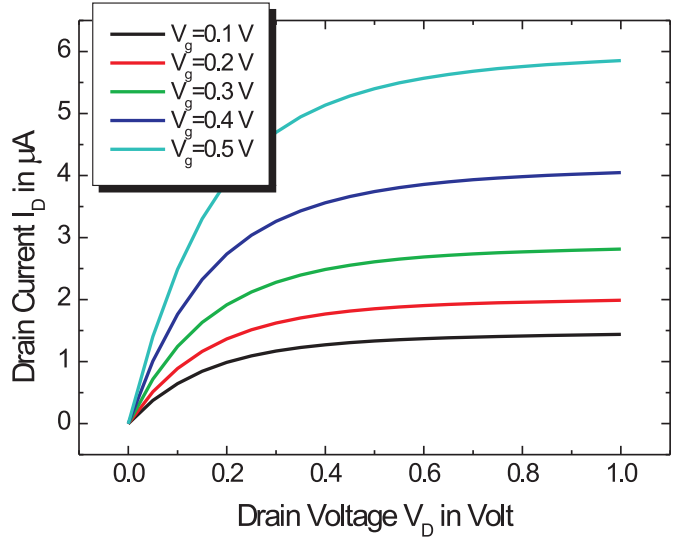}} 
 \caption{Drain current, $I_D$, as a function of drain voltage, $V_D$, for different gate voltages, $V_G = V_{G1} = V_{G2}$.}
 \label{FET}
\end{figure}

For the measurements in Figure \ref{CONF} the two gate contacts were connected together and the gate current was kept constant at $I_G = 260\,\mu$A. At zero drain voltage the drain current, $I_D$, is initially negative and the channel operates as a light-emitting diode at both the source and drain side. However, as the drain voltage, $V_D$, is increased to positive values, the drain current, $I_D$, changes sign and eventually saturates. This situation will occur when the depletion region on the drain side expands along the $z$-direction across the whole $i$-GaAs channel. This behavior is characteristic for an FET device and is shown in more detail in Figure \ref{FET}. Therefore, when $V_D$ is used to confine the recombination region the current $I_D$ into the drain contact will saturate at a constant value while the recombination zone is squeezed near the source side.

\section{Device simulation}
\label{DevSimChap}

The following numerical analysis is intended to illustrate the principle of electrostatic confinement of both carrier types into the same region. The simulation calculates the concentration of electrons in the conduction band, $n_c$, and holes in the valence band, $p_v$, as a function of $y$ and $z$ using Poisson's equation $\Delta \phi = \frac{\rho(y, z)}{\epsilon}.$ In this way the electrostatic potential, $\phi$, relates to the total space charge density, $\rho$, where $\epsilon$ is the local permittivity. The local space charge density is the sum of contributions from all mobile and fixed charges, including electrons, holes and ionized impurities with impurity concentration due to donors $N_D^+$ and acceptors $N_A^-$, as well as charge due to traps and defects $Q_T$. Thus 
\[
 \rho = e (n_c - p_v - N_D^+ + N_A^-) + Q_T,  
\]
where $e$ is the electron charge. 

The drift-diffusion model is assumed to model the non-equilibrium situation where the existence of both an electric field $E=\nabla \phi$ and a carrier density gradient $\nabla n_c$ and $\nabla p_v$ lead to finite electron and hole currents $J_e$ and $J_h$. The carrier current density can be written as:
\begin{eqnarray}
  J_e & = & - \mu_n n_c E - D_n \nabla n_c, \nonumber \\ 
  J_h & = & \mu_p p_v E - D_p \nabla p_v, \nonumber
\end{eqnarray}
where $\mu_n$ and $\mu_p$ are the electron and hole mobilities. The currents $J_e$ and $J_h$ together with several recombination processes described by $\left(\frac{d n_c}{d t}\right)_{r}$ and $\left(\frac{d n_c}{d t}\right)_{r}$ determine the change in the carrier densities via the continuity equation:
\sffamily
\begin{eqnarray}
\label{eqnCont}
  \frac{\partial n_c}{\partial t} & = & \left(\frac{d n_c}{d
      t}\right)_{r} - {\rm div} J_e,\nonumber \\ \nonumber
  \frac{\partial p_v}{\partial t} & = & \left(\frac{d p_v}{d
      t}\right)_{r} -{\rm div} J_h.
\end{eqnarray}
\normalfont
Possible recombination processes include photon transitions, phonon transitions and surface recombination. 

\begin{figure}
 \centerline{\includegraphics[scale=1.0]{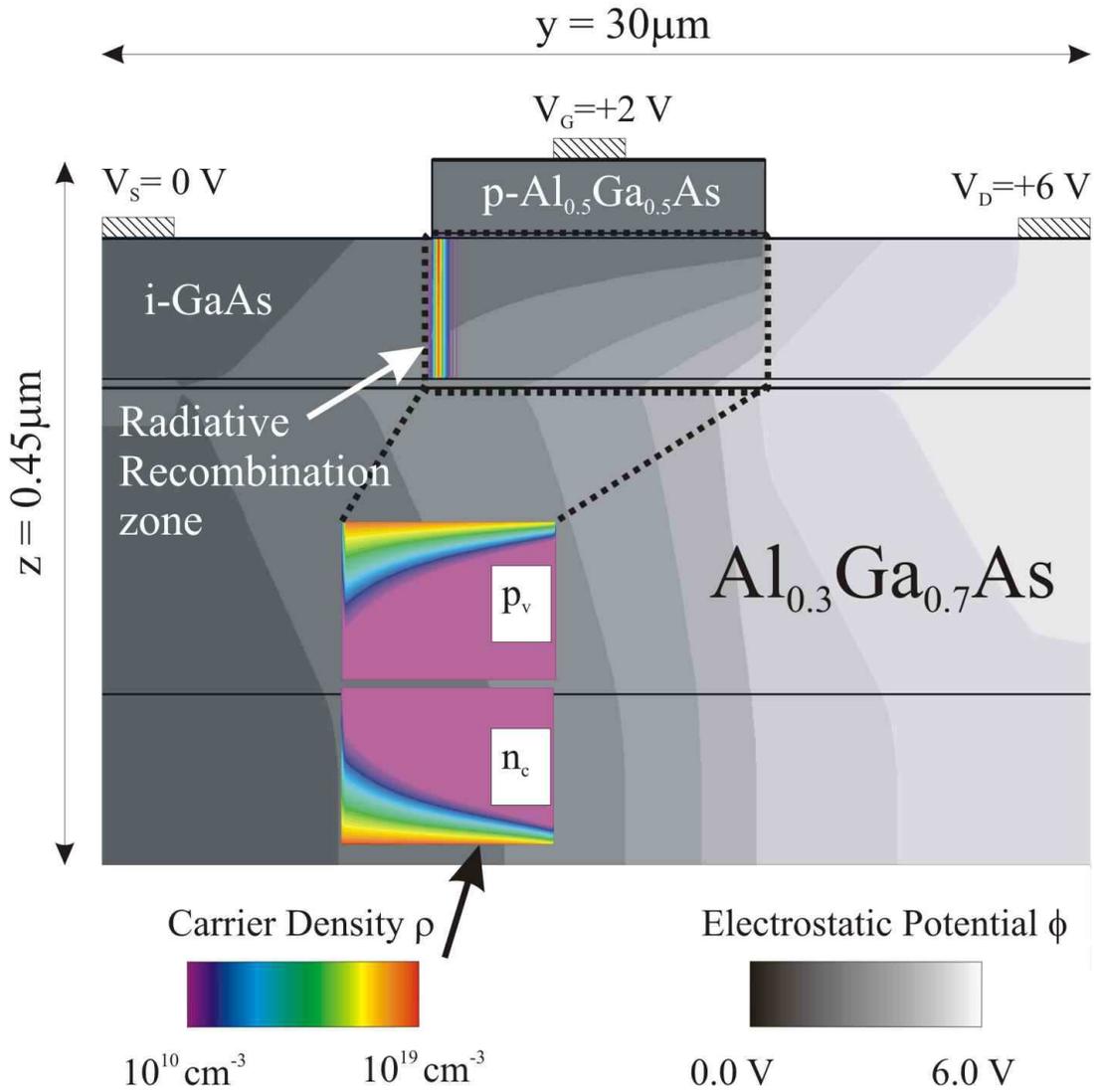}} 
 \caption{Result of two-dimensional device simulation using device simulator ATLAS. The distribution of the electrostatic potential $\phi$ is shown in gray scale. The calculated radiative recombination region near the left step edge is shown by colored lines. The colored insets show the electron (lower inset) and hole (upper inset) density across the area enclosed by the dashed box in the $i$-GaAs channel.}
 \label{Simulation}
\end{figure}

These equations are solved numerically using the device simulator ATLAS from Silvaco on a two-dimensional grid. The problem is transfered into two dimensions by considering a $30 \mu m$ by $0.5 \mu m$ large cross-section perpendicular to the $x$-axis. Three ohmic contacts are placed on top of the surface exposed to vacuum as shown in Figure \ref{Simulation}. Along the outer edges of devices, homogeneous (reflecting) Neumann boundary conditions are imposed so that current only flows out of the device through the contacts. In the absence of surface charge along such edges, the normal electric field component becomes zero. However, surface charges are assumed on the surfaces exposed to vacuum to pin the Fermi-level at midgap. In this way realistic values of the electrostatic potential distribution, $\phi(y,z)$, may be obtained as confirmed by secondary electron microscopy in Ref. \cite{kaestner}.

The gray area in Figure \ref{Simulation} shows the electrostatic potential distribution, $\phi(y,z)$, over the cross-section of the device. The colored insets show the electron (lower image) and hole (upper image) concentration of the region inside the marked dashed box. The potential drop at the right step-edge inside the $i$-GaAs region is much larger than across the left step-edge. Consequently, electrons and holes are repelled from the right step-edge, as shown in the colored insets of Figure \ref{Simulation}. However, holes cannot be transported much beyond the left step-edge because of the potential barrier formed by the fixed space charges above and below the $i$-GaAs region. In this way the recombination region is located close to the left step-edge. The numerically obtained recombination zone has been marked using colored lines. As observed experimentally in the previous section its size is small compared to its distance to the contact.

\section{Conclusion}

We have presented a four-terminal light-emitting transistor implemented in a GaAs/AlGaAs heterostructure. It was shown both experimentally and numerically that in this device electrons and holes can be directed to the same area by drain and gate voltages, defining a recombination zone tunable in size and position. Based on this architecture devices may be designed suitable for optically probing low-dimensional transport as well as for the generation of non-classical light. The existence of two gate contacts may be used to add an additional functionality compared to traditional three terminal transistors in that the gate may carry a current itself. In addition it is possible to fabricate self aligned gates next to an EL probing device.

Many applications of this structure require low temperature such that the presented material system was designed for temperatures between 5-10K. However, the wafer composition can be adjusted to obtain similar characteristics at room temperature. Finally, the structure design is general enough to transfer it to other direct bandgap material systems, such as GaN or InSb.

\end{document}